\documentclass[letterpaper, 10 pt, conference]{ieeeconf}  

\IEEEoverridecommandlockouts                              

\usepackage{graphics} 
\usepackage{epsfig} 
\usepackage{amsbsy}
\usepackage{dsfont}
\usepackage{mathtools}
\usepackage{bm}
\usepackage{subfiles}
\usepackage[noend,ruled,linesnumbered]{algorithm2e}
\usepackage{caption}
\usepackage{subcaption}
\usepackage{cite}
\usepackage{amsmath, amssymb}
\usepackage{bbm}
\usepackage{cite}
\usepackage{hyperref}
\DeclareCaptionFont{mysize}{\fontsize{8}{9.6}\selectfont}
\usepackage{adjustbox}
\newtheorem{assumption}{\textbf{Assumption}}

\newtheorem{lemma}{\textbf{Lemma}}
\newtheorem{remark}{\textbf{Remark}}
\newtheorem{problem}{\textbf{Problem}}
\newtheorem{theorem}{\textbf{Theorem}}
\newtheorem{definition}{\textbf{Definition}}

\DeclareMathOperator*{\argmin}{arg\,min}

\usepackage{lipsum}

\makeatletter
\newcommand*{\centerfloat}{%
  \parindent \z@
  \leftskip \z@ \@plus 1fil \@minus \textwidth
  \rightskip\leftskip
  \parfillskip \z@skip}
\makeatother

\title{\LARGE \bf
Deep Equivariant Multi-Agent Control Barrier Functions
}

\author{Nikolaos Bousias$^{1}$, Lars Lindemann$^{2}$ and George Pappas$^{1}$
\thanks{*supported by the ARL grant DCIST CRA W911NF-17-
2-0181.}
\thanks{$^{1}$GRASP Lab, Department of Electrical \& Systems Engineering, University of Pennsylvania, Philadelphia PA, 19104, USA. $^{2}$Department of Computer Science, University of Southern California, Los Angeles, California
        {\tt\small nbousias@seas.upenn.edu}}%
}

\begin{document}

\maketitle
\thispagestyle{empty}

\pagestyle{empty}

\begin{abstract}
With multi-agent systems increasingly deployed autonomously at scale in complex environments, ensuring safety of the data-driven policies is critical. Control Barrier Functions have emerged as an effective tool for enforcing safety constraints, yet existing learning-based methods often lack in scalability, generalization and sampling efficiency as they overlook inherent geometric structures of the system. 
To address this gap, we introduce symmetries-infused distributed Control Barrier Functions, enforcing  the satisfaction of intrinsic symmetries on learnable graph-based safety certificates. We theoretically motivate the need for equivariant parametrization of CBFs and policies, and propose a simple, yet efficient and adaptable methodology for constructing such equivariant group-modular networks via the compatible group actions. This approach encodes safety constraints in a distributed data-efficient manner, enabling zero-shot generalization to larger and denser swarms. 
Through extensive simulations on multi-robot navigation tasks, we demonstrate that our method outperforms state-of-the-art baselines in terms of safety, scalability, and task success rates, highlighting the importance of embedding symmetries in safe distributed neural policies.

\end{abstract}

\section{Introduction}
Multi-agent systems are increasingly being deployed autonomously in complex environments as a means to complete complex tasks efficiently and with redundancy. An ubiquitous, critical requirement for mission success, though, is to guarantee safety in terms of collision avoidance in a scalable and computationally efficient manner. 
Common approaches like MILP \cite{10529204}, H-J Reachability analysis \cite{7798509}, sampling-based planning \cite{7911207} and MPC \cite{8613928} are computationally intractable for larger swarms and high order dynamics, while Multi-Agent RL \cite{7989037, zhang2024scalable} offers no safety guarantees as competing rewards trade between conflicting performance and safety incentives. 

Control Barrier Functions have emerged as a powerful tool to guarantee forward invariance in the perceived safe space \cite{ames2016control}. Since such centralized methods still struggled with scalability \cite{glotfelter2017nonsmooth}, distributed versions were constructed \cite{9642050,8718798,9992744}. Though computationally efficient and scalable, the latter proved overly conservative, restricting the task performance of the system, and particularly tedious to synthesize for multi-robot systems of highly non-linear dynamics. To address these issues, data-driven \cite{10.1109/CDC42340.2020.9303785} distributed methods \cite{mdcbf2021,zhang2023gcbf,zhang2025gcbf+,gao2024provably} were introduced with a neural network of appropriate structure approximating the CBF. In \cite{mdcbf2021}, the authors introduced a framework for jointly learning scalable decentralized CBFs and policies. However, it failed to account for the dynamic geometric topology of the problem or distinguish between cooperative agents and uncooperative obstacles, leading in conservative behaviors or safety breaches. To that end, \cite{gao2024provably} exploited the dynamic graph representation of the problem, harnessing the demonstrated scalability of Graph Neural Networks and their centralized-training, decentralized-execution nature to limit communication and instead base inter-agent cooperation for satisfaction of safety constraints on policy homogeneity \cite{zhang2025gcbf+}. 

Learning multi-agent policies, however, is a sampling inefficient process that requires numerous demonstrated interactions. Most crucially, deploying black-box learning-based methods in safety critical applications requires some generalization assurances to out of training distribution, but at least predictable, data.
One promising avenue to address these challenges lies in leveraging the inherent geometric structure of the safety constrained control problem. By identifying existing symmetries and embedding them as inductive bias in the learned parametrizations of the CBF and policy functions, one shrinks the hypothesis class of models to only those that satisfy the problem structure, supported by the data. This reduces the required parameters of the network \cite{bekkers2024fast}(and thus the required data to fit it on) and offers some predictability to samples that are out of the dataset but still a transformation of examples from it. 
The existence of geometric symmetries in safety certificates and policies, morphs into equivalence of state-control pairs under a compatible group transformation, meaning that the networks only needs to learn one mapping for each equivalence class, rather than separately learning the same behavior for all symmetrically related pairs. If these symmetries are local, they will exist independently of the size of the swarm. Symmetry-enhanced models have shown promise in improving learning efficiency and scalability in multi-agent policies \cite{vanderpol2022multiagentmdphomomorphicnetworks, mcclellan2024boostingsampleefficiencygeneralization}. An often leveraged omnipresent symmetry in homogenous MASs is permutation equivariance, i.e. indexing is interchangeable \cite{zhang2023gcbf}. Nevertheless, there is a significant gap in leveraging other geometric symmetries for distributed safe control, that this paper attempts to bridge. Furthermore, the existing models are commonly based on equivariant operations on message-passing networks \cite{pmlr-v139-satorras21a,fuchs2020se3transformers} that are not transferable to systems with different symmetries. To alleviate this issue, we propose a simple, yet fast and modular, approach that can be adapted for a variety of symmetries and architectures.


\noindent\textbf{Contributions}
The aforementioned literature, permutation equivariance aside, largely ignores the geometric structure of the distributed safe control problem. This paper presents a novel framework for leveraging the intrinsic symmetries exhibited by the system to enhance safety generalization and scalability of the learned CBF-policy. Our contributions are outlined below:
\begin{itemize}
    \item  We introduce the symmetry-infused multi-agent control barrier functions, that guarantee forward invariance in the safe set, and provide intuition as to why exploiting symmetries could enhance data-driven distributed CBFs and policy generalization. We, further, provide a formalization of the conditions under which optimal safe distributed policies are equivariant functions to be approximated by equivariant networks.
    \item We propose methodology for constructing graph-processing networks that respect exhibited symmetries of the system. This method provides a simple, yet efficient, way to ensure symmetry satisfaction and is by design symmetry-modular, as any group compatible with the manifold of the robot may be used, in contrast to the vast majority of networks in the literature.
\end{itemize}
To substantiate our claims, we provide experiments, demonstrating significant improvements in scalability, generalization, and safety compared to state-of-the-art methods. 

\section{Preliminaries}

\subsection{Group Theory \& Equivariant Functions}
A group $(G,\cdot)$ is a set $G$ equipped with an operator $\cdot:G\times G \rightarrow G$ that satisfies the properties of \emph{Identity:} $\exists e \in G$ such that $e \cdot g = g \cdot e = e$; \emph{Associativity:} $\forall g,h,f \in G,\, g\cdot (h \cdot f) = (g \cdot h) \cdot f$; \emph{Inverse:} $\forall g \in G,\, \exists g^{-1}$ such that $g^{-1} \cdot g = g \cdot g^{-1} = e$. Additional to its structure we can define the way that the group elements act on a space $X$ via a group action:
\begin{definition}\label{definitioin:group_action}
A map $\phi_g: X\to X$ is called an action of group element $g\in G$ on $X$ if for $e$ is the identity element $\phi_e(x)=x$ for all $x\in X$ and $\phi_g\circ \phi_h=\phi_{g\cdot h}$ for all $g,h\in G$.
\end{definition}

Note here that a group action on a given space $X$ allow us to group different elements of $X$ in sets of orbits. More precisely given a group action $\phi_*$ an orbit of a element $x\in X$ is the set $\mathcal{O}_x^{\phi_*}=\left\{\phi_g(x)|g\in G\right\}$. 
\noindent In many application we require functions that respect the structure of a  group acting on their domain and codomain. 
We refer to these functions as equivariant and we formally define them as follow:
\begin{definition}
Given a group $G$ and corresponding group actions $\phi_g:X\to X$, $\psi_g:X\to X$ for $g\in G$  a function $f:X\to Y$ is said to be $(G,\phi_*,\psi_*)$ equivariant if and only if $\psi_g(f(x))=f\left(\phi_g(x)\right)\,, \forall x\in X, g\in G$.
\end{definition}

\subsection{Notation}
    A continuous function $\alpha:\mathbb{R}\rightarrow \mathbb{R}$ is an extended class-$\mathcal{K}$ function if it is strictly increasing and $\alpha(0)=0$. Let $[\cdot]_+=\max\{0,\cdot\}$ and $\bar{\zeta}$ the complement of a set $\zeta$.
    Let $\mathcal{X}$ be a smooth manifold and $T_x\mathcal{X}$ the tangent space at an arbitrary $x\in \mathcal{X}$. A smooth vector field is a smooth map $f:\mathcal{X}\rightarrow T\mathcal{X}$ with $f(x)\in T_x\mathcal{X}$. The set of smooth vector fields on a manifold $\mathcal{X}$, denoted $\mathfrak{X}(\mathcal{X})$, is a linear infinite dimensional vector space. Let $G$ be a d-dimension real Lie group, with identity element $e$. 
A group action is \textit{free} if $\forall x \in \mathcal{X},\phi_g(x)=x \Leftrightarrow g=e $ and \textit{transitive} if $\forall x,y \in \mathcal{X} \,,\,\exists g\in G$ such that $\phi_g(x)=y$ (i.e. the nonlinear smooth projections $\phi_x(g):=\phi(g,x)$ are surjective). 
    A \textit{homogeneous} space is a smooth manifold $\mathcal{X}$ that admits a transitive group action $\phi: G \times \mathcal{X} \rightarrow \mathcal{X}$ and the Lie group $G$ is, then, called the \textit{symmetry} of $\mathcal{X}$. The group torsor $\mathfrak{G}$ of a Lie group $G$ is defined as the underlying manifold of $G$ without the group structure, allowing for identification of the torsor elements by the group elements, denoted $\chi\in \mathfrak{G} \simeq g \in G$, and inheriting the free and transitive group action $\phi$ induced by the group operator, i.e. for $\hat{g}\in G$ and $\chi\in \mathfrak{G} \simeq g \in G$ it stands that $\phi(\hat{g},\chi) \simeq \hat{g}\cdot g$. Crucially, a manifold serving as a torsor for multiple Lie groups may admit multiple symmetries.

\subsection{Problem Formulation}
Consider a homogeneous multi-robot dynamical system, comprising of $N$ autonomous robots indexed $i\in \{1,\dots,N\} \equiv I_N$. Let $\mathcal{X}$ be a smooth manifold and $\mathcal{U}$ a finite dimensional input space. The agents are described by control affine dynamics $\dot{x}_i(t) = f(x_i(t),u_i(t)),\quad \forall i\in I_N$,
where $u_i\in \mathcal{U},x_i\in\mathcal{X}$ the input and state of the robot respectively and $f:\mathcal{X}\times \mathcal{U}\rightarrow \mathfrak{X}(\mathcal{X})$ locally $\epsilon$-Lipschitz continuous. Consider a Lie group $G$ and a smooth transitive group action $\phi:G \times \mathcal{X}\rightarrow \mathcal{X}$. Let $p_i(t) \in \mathcal{P} \subset \mathcal{X},\forall i \in I_N$ denote the position of the robots, with $x_i(t) = [p_i(t),q_i(t)], q_i(t)\in \mathcal{X} \setminus \mathcal{P}$. 
The task is to learn a distributed control policy (formally defined below) that safely directs the robots to desired states $\hat{x}_i\in \mathcal{X},\forall i \in I_N$. To navigate around the obstacle-cluttered environment, the robots are equipped with sensors, acquiring $w$  observations in the form of LiDAR rays $y_{j}^{i}(t)\in \mathcal{P},j\in W\equiv\{2N,...,2N+w\}$ that provide measurements on obstacles within a range $R\in \mathbb{R}^+$. Let $\bar{y}_j^i(t)=[y_j^i(t),q_0]\in \mathcal{X},q_0\in \mathcal{X}\setminus \mathcal{P}$ denote the padded observations. 

\subsubsection{Graph Representation of Multi-robot System}\label{intro:graph_representation}
We assume that the robots are equipped with communication capabilities with a range $\hat{R}>R$. Let $\mathcal{N}^{\hat{R}}_i:= \{j\in I_N \setminus\{i\} \,:\, ||p_i-p_j||\leq \hat{R} \}\,,\, \forall i \in I_N$ denote the neighbourhoods, thus giving rise to a graph representation of the multi-robot system $\mathcal{G}_t=\{V
_{\mathcal{G}_t},\mathcal{E}_{\mathcal{G}_t}\}$, with nodes $V
_{\mathcal{G}_t}=\{x_i(t)\,,\, i\in I_N\}$ and edges representing the flow of information $\mathcal{E}_{\mathcal{G}_t}=\{ (i,j)\,,\, : \forall i\in I_N,j\in \mathcal{N}^{\hat{R}}_i \}$. Similarly to \cite{INFOMARL}, we extend the graph representation to incorporate target and obstacle states, i.e. $\hat{\mathcal{G}}_t=\{V
_{\hat{\mathcal{G}}_t},\mathcal{E}_{\hat{\mathcal{G}}_t}\}$, with nodes $V
_{\hat{\mathcal{G}}_t}=V
_{\mathcal{G}_t} \cup \{\bar{y}_j^i(t)|\forall i \in I_N,j \in W\} \cup \{\hat{x}_i|\forall i \in I_N\}$ and edges representing the flow of information $\mathcal{E}_{\hat{\mathcal{G}}_t}=\mathcal{E}_{\mathcal{G}_t} \cup \{(i,j)| \forall i \in I_N, j \in W\} \cup \mathcal{E}_{*} $ for $\mathcal{E}_{*}=\{(i,l)|\forall i\in I_N,l \in I_N+N\}$. Assuming that a submanifold $\Bar{X} \subseteq\mathcal{X}$ forms the torsor $\mathfrak{G}$ of the Lie group $G$ and that $\mathcal{X}\setminus \Bar{X}$ is compatible with $G$, then every robot inherits a group representation element $g_i\,,\forall i\in I_N$ and the node attributes of the graph become $V_{\hat{\mathcal{G}}_t}=\{(g_i(t),v_i(t))| \forall i \in |V_{\hat{\mathcal{G}}_t}|\}$, thus providing the geometric structure of the graph. The node features are padded with an encoding that distinguishes robot, object and target nodes. For convenience we denote $\hat{\mathcal{G}}_t^{i}$ the subgraph of the $G$-augmented $\hat{\mathcal{G}}_t$ that forms the immediate neighborhood of agent $i$.

\subsubsection{Safety constrained distributed control}\label{intro:problem_statement}
We denote $x^{\bar{I}_N}(t)=\oplus_{i\in \bar{I}_N} x_i(t) \in \mathcal{X}^N, u^{\bar{I}_N}(t)=\oplus_{i\in \bar{I}_N} u_i(t)\in \mathcal{U}^N,\bar{I}_N \subseteq I_N$ the $\bar{I}_N$-centralized state/input with respective dynamics $\dot{x}^{\bar{I}_N}(t) = f^{\bar{I}_N}(x^{\bar{I}_N}(t),u^{\bar{I}_N}(t)) = \oplus_{i\in \bar{I}_N} f(x_i(t),u_i(t))$. 
Consider the safe set $S_{N}\subseteq \mathcal{X}^N$, the set of all states of the MAS that satisfy the collision avoidance, with respect to robots and obstacles, specification, i.e. $S_{N}^{r\in [0,R)} := \{ x^{I_N}\in \mathcal{X}^N | \, || p_i-p_j|| >r, \forall (i,j) \in \mathcal{E}_{\hat{\mathcal{G}}_t} \setminus \mathcal{E}_{*},i\neq j\} $.
The problem is to design a distributed control policy $u_i(t)=\pi(\hat{\mathcal{G}}_t^{i}),\forall i \in I_N$ that safely guides the robots to their respective target states, i.e. satisfies liveness ($\inf_{t} ||p_i(t)-\hat{p}_i||=0, \forall i \in I_N$) and safety constraints ($x^{I_N}(t)\in S^r_N$).  
\begin{problem}\label{problem:1}
Given a swarm of $N$ robots and compatible target states $\hat{x}_i,\forall i \in I_N$, solve
    \begin{align}\label{eq:prob}
    u^{I_N,*}_{0:T}=& \argmin_{u^{I_N}_{0:T}} \int_{0}^T \mathcal{T}_N(x^{I_N}(\tau)|\hat{x}^{I_N}) d\tau \nonumber\\
    s.t.\quad& \dot{x}_i(t)=f(x_i(t),u_i(t))\in \mathfrak{X}(\mathcal{X})\,,\, i\in I_N \nonumber\\
    &u_i(t) = \pi(\hat{\mathcal{G}}_t^{i}) \in \mathcal{U},\, x^{I_N}(t)\in S^r_N
    \end{align}  
where $\mathcal{T}_N(x^{I_N}(\tau)) = \sum_{\i\in I_N} ||p_i(t)-\hat{p}_i||$.
\end{problem}
Problem \ref{problem:1} is challenging since the safe set $S_N$ is not available and decomposed to individual agent-centric sets, and the distributed policy requires generating agent actions based on local neighborhood information. 
\begin{assumption}
    Parameter $R$ is sufficiently large to ensure Problem \ref{problem:1} is always feasible once obstacles are discovered.
\end{assumption}

\section{Safety Certificates for Multi-agent Systems}
Control barrier functions are a common method for verifying that the state of dynamical system is forward invariant in the safe set \cite{8796030}.
\begin{definition}
The multi-agent system represented by the graph structure $\hat{\mathcal{G}}_t$ is safe at time $t$, i.e. $x^{I_N}(t) \in S^r_N$ if and only if all individual agent-centric neighborhoods represented by the subgraphs $\hat{\mathcal{G}}^i_t$ are safe at time $t$, i.e. $x^{\mathcal{N}^{\hat{R}}_i \cup \{i\}}(t)\in S^r_{N_,i}:= \{ x^{{\mathcal{N}^{\hat{R}}_i}}\in \mathcal{X}^{|\mathcal{N}_i|+1} | \, || p_i-p_j|| >r, \forall j \in \mathcal{E}_{\hat{\mathcal{G}}^i_t} \setminus \mathcal{E}_{*}\}  \,,\,\forall i \in I_N$.
    \label{def:1}
\end{definition}
Definition \ref{def:1} allows for a distributed viewpoint of the notion of multi-agent safety via local safety certificates.

\subsection{Graph-based Control Barrier Functions}\label{section:graph_cbf}
Leveraging Definition \ref{def:1}, the safety specification of a MAS is decomposed to the safety specification of each individual node of the graph, that is the graph representation is safe if the all 1-step subgraphs are safe, thus providing for a decentralized safety certificates. 
\begin{assumption}
    The safety of node $i\in I_N$ is only affected by nodes in its $R$-neighborhood $j\in \mathcal{N}^R_i\subseteq \mathcal{N}^{\hat{R}}_i$.
\label{ass:2}
\end{assumption}
\begin{definition} \label{def:macbf}
    Consider a continuously differentiable function $h:\mathcal{X}^{|\mathcal{N}|+1}\rightarrow \mathbb{R}$ and the graph representation $\hat{\mathcal{G}}_t$ of section \ref{intro:graph_representation}. Let Assumption \ref{ass:2} stand and $C_N \subset S^r_N$. Then $h$ is a valid distributed CBF if there exists a locally Lipschitz continuous extended class-$\mathcal{K}$ function $\alpha:\mathbb{R}\rightarrow \mathbb{R}$ such that 
    \begin{align}\label{cbf_contraint_def}
        \sup_{u^{\mathcal{N}^{\hat{R}_i\cup \{i\} } }\in \mathcal{U}^{|\mathcal{N}^{\hat{R}}_i|+1}} \Big[ 
        &\sum_{j\in \mathcal{N}^{\hat{R}_i \cup \{i\}} } \langle \nabla_{x_j} h(x^{\mathcal{N}^{\hat{R}}_i\cup \{i\}}) , f(x_j,u_j)
        \rangle \Big] \nonumber\\
        &\geq -\alpha(h(x^{\mathcal{N}^{\hat{R}}_i\cup \{i\}}))
    \end{align}
\end{definition}
where $\mathcal{B}_i:=\{ x^{\mathcal{N}_i \cup \{i\}} \in \mathcal{X}^{|\mathcal{N}_i|+1} | h(x^{x^{\mathcal{N}_i \cup \{i\}}}) \geq 0\}$ and $\forall N \in \mathbb{N}, \mathcal{C}_N=\cap_{i\in I_N}\{x^{I_N}\in \mathcal{X}^N| x^{\mathcal{N}_i \cup \{i\}} \in \mathcal{B}_i\} \subset \cap_{\forall i \in I_N}S^r_{N_,i} \subseteq S^r_N $. 
\begin{remark}
    For a valid distributed CBF $h$ of Definition \ref{def:macbf}, Assumption \ref{ass:2} amounts to $\nabla_{x_j} h(x^{\mathcal{N}^{\hat{R}}_i \cup \{i\} })=0,\forall j \in \mathcal{N}^{\hat{R}}_i \setminus \mathcal{N}^R_i $ and $h(x^{\mathcal{N}^{\hat{R}}_i\cup \{i\}})=h(x^{\mathcal{N}^{R}_i\cup \{i\}})$. 
\end{remark}
\begin{remark}
    Although node attributes evolve on a continuous manifold $\mathcal{X}^{|\mathcal{N}^{\hat{R}}_i|}$, $x^{\mathcal{N}^{\hat{R}}_i\cup \{i\}}$ has discontinuities due to changes in the topology of $\hat{\mathcal{G}}^i_t$. Still, a valid distributed CBF $h$ of Definition \ref{def:macbf}, is continuously differentiable \cite{zhang2025gcbf+}.
\end{remark}
\begin{lemma}\label{lemma:cbf}
    Assume a valid CBF $h:\mathcal{X}^{|\mathcal{N}|+1}\rightarrow \mathbb{R}$, with $C_N \subseteq \mathcal{D}\subset S^r_N$. For a locally Lipscitz continuous policy $\pi:\mathcal{X}^{N} \rightarrow \mathcal{A}^\text{CBF}$ and any swarm size $N\in \mathbb{N}$, if $x^{I_N}(0)\in \mathcal{C}_N$ then $x^{I_N}(t)\in \mathcal{C}_N \subset S^r_N,\forall t$ \cite{zhang2025gcbf+}. If $C_N$ is compact, it additionally follows that $C_N$ is forward invariant and $x^{I_N}(t\rightarrow \infty) \rightarrow \mu \in \mathcal{C}_N$ asymptotically stable if $x^{I_N}(0)\in \bar{\mathcal{C}_N}\cap \mathcal{D}$ \cite{10.1109/CDC42340.2020.9303785,ames2016control,XU201554}.
\end{lemma}
Note that $h(x^{\mathcal{N}^{\hat{R}}_i \cup \{i\} })= 0, \forall x^{\mathcal{N}^{\hat{R}}_i \cup \{i\} } \in \partial(\mathcal{C}_{N,i})$ is not required when using the Comparison Lemma instead of Nagumo’s theorem \cite{8796030}.

\subsection{Learning Safe Distributed Policy}\label{section:safe_distributed_policy}
Applying Lemma \ref{lemma:cbf} in Problem \ref{problem:1} leads to:
\begin{align}\label{eq:prob2}
    \min&_{u_i(0:T),i\in I_N} \int_{0}^T \sum_{i\in I_N} \mathcal{T}(x_i(\tau)|\hat{x}_i ) d\tau \\
    s.t.\quad& \dot{x}_i(t)=f(x_i(t),u_i(t))\in \mathfrak{X}(\mathcal{X}) \\
    &u_i(t) = \pi(x^{\mathcal{N}^{\hat{R}}_i\cup \{i\}}(t)) \in \mathcal{U} \\
    & \sum_{j\in \mathcal{N}^{\hat{R}_i \cup \{i\}} } \langle \nabla_{x_j(t)} h(x^{\mathcal{N}^{\hat{R}}_i\cup \{i\}}(t)) , f(x_j(t),u_j(t)) \rangle \nonumber\\
    & \quad\quad\quad  \geq -\alpha(h(x^{\mathcal{N}^{\hat{R}}_i\cup \{i\}}(t)))\,,\, \forall i\in I_N \label{cbf_constraint}
    \end{align}  
Though the objective function only depends on $x_i(t)$, the CBF constraints depend on $u_j(t),j\in\mathcal{N}^{\hat{R}_i\cup \{i\}}$, leading to a centralized optimization problem, meaning that to guarantee the safety of an agent, the cooperation of neighboring agents is required. Solving a centralized non-convex optimization problem online, however, is tedious and computationally intractable, particularly for large swarm sizes. To tackle this, we adopt an on-policy scheme to jointly learn distributed candidate CBF $h_\phi$ and distributed collision-avoidance control policy $\pi_\theta$ in a centralized-training, distributed-inference setup. With the graph representation introduced in Section \ref{intro:graph_representation}, we can parametrize $h_\phi,\pi_\theta$ with a GNN-based architecture. This approach is well-suited for processing signals on graphs, handles varying neighborhood sizes, and exhibits permutation equivariance \cite{tzes2023graph}, which significantly improves sample efficiency by enabling robots to share experiences. 

Similarly to \cite{gao2024provably}, to solve the problem defined by equations \ref{eq:prob2}-\ref{cbf_constraint}, a surrogate hierarchical approach for the liveness specification is employed. Instead of optimizing the target-reaching objective $\int_{0}^T \sum_{i\in I_N}\mathcal{T}(x_i(\tau)|\hat{x}_i (\tau))$, liveness of the policy could result from imitating a nominal controller $u_{i,\text{nom}}=\pi_\text{nom}(x_i(t),\hat{x}_i)$. As $\pi_\text{nom}$ lacks safety context, the policy should leverage as reference the solution of the min-norm, safety constrained problem:
\begin{align}\label{problem:3}
    &\min_{u_i\in \mathcal{U},\forall i\in I_N}  ||u_i-u_{i,\text{nom}}||^\kappa \nonumber\\
    &s.t. \quad\quad \text{Equation (\ref{cbf_constraint})}
\end{align}
For control affine dynamics and convex $\mathcal{U}$, \ref{problem:3} is a QP with optimal solution $\pi_\text{QP}:\mathcal{X}^N\rightarrow \mathcal{U}^N$. 

\noindent\textbf{Loss function}: Consider an on-policy collected dataset comprising of safe control invariant demonstrations $\mathcal{D}_{S,i}=\{ (\hat{\mathcal{G}}^i_{\{0:k\}}, u^{\mathcal{N}^{\hat{R}_i \cup \{i\}}}) | x^{\mathcal{N}^{\hat{R}_i \cup \{i\}}} \in S_{N,i}\}$ and unsafe ones $\mathcal{D}_{U,i}$, respectively and $\mathcal{D}_i=\mathcal{D}_{S,i} \cup \mathcal{D}_{U,i}$. Following \cite{mdcbf2021}, parameters $\theta,\phi$ are trained with the loss function
\begin{align}
    \mathcal{L}_{\theta,\phi} &:= \sum_{i\in I_N} \Bigg[ \eta_c||\pi_\theta(x^{\mathcal{N}^{\hat{R}}_i\cup \{i\}}) - \pi_\text{QP}(x^{I_N})|| \,+ \nonumber\\
    &\eta_d \sum_{\mathcal{G}\in \mathcal{D}_i} \bigg[ \gamma - \sum_{j\in \mathcal{N}^{\hat{R}_i \cup \{i\}} } \langle \nabla_{x_j} h_{\phi}(x^{\mathcal{N}^{\hat{R}}_i\cup \{i\}}) , f(x_j,u_j) \rangle \nonumber\\
    & -\alpha(h_{\phi}(x^{\mathcal{N}^{\hat{R}}_i\cup \{i\}}))  \bigg]_{+} + \sum_{\mathcal{G}\in \mathcal{D}_{S,i}} \bigg[\gamma - h_{\phi}(x^{\mathcal{N}^{\hat{R}}_i\cup \{i\}}) \bigg]_{+} \nonumber\\
    & +\sum_{\mathcal{G}\in \mathcal{D}_{U,i}} \bigg[\gamma + h_{\phi}(x^{\mathcal{N}^{\hat{R}}_i\cup \{i\}}) \bigg]_{+}
\end{align}

During training, we collect and label data in an on-policy method (Section \ref{sec:experiments}) and  use $h_\phi,\pi_\text{nom}$ to solve the centralized QP-CBF, aquiring $\pi_\text{QP}$, similarly to \cite{zhang2025gcbf+}. The loss function $\mathcal{L}$ is then computed and the errors backpropagated to update $\pi_\theta,h_\phi$. Notice that, during training only, from equation \ref{cbf_constraint}, the safety of agent $i$ depends also on $u^{\mathcal{N}^{\hat{R}_i}}$, thus via the backpropagation, $\pi_\theta(\hat{\mathcal{G}}^j_t)$ is further affected by $\pi_\theta(\hat{\mathcal{G}}^j_t)$, illustrating experience sharing between neighbors.

\section{Symmetries-infused Safe Policies}
In this section we introduce the group invariant multi-agent CBFs and provide intuition as to why leveraging symmetries could enhance the learned CBF and policy generalization. We, then, introduce the Equivariant Graphormer, a novel group-modular graph-based architecture that respects group symmetries via the group induced canonicalizing actions.
\begin{definition}\label{def:equiv_dynamics}
    Consider a real Lie Group $G$, a smooth manifold with group properties satisfying differentiability of group operations. The robot dynamics are $G$-equivariant if, for transitive actions induced by elements of $G$ on a vector field $\mathcal{X}$, $\phi:\mathcal{X}\times G \rightarrow \mathcal{X}$ and $ \psi:\mathcal{U}\times G \rightarrow \mathcal{U}$, satisfies $d\phi_gf(x_i(t),u_i(t)) = f(\phi_g(x_i(t)), \psi_g(u_i(t)),\,\forall g \in G $, where $d\phi: \mathfrak{X}(\mathcal{X}) \times G\rightarrow \mathfrak{X}(\mathcal{X})$ the differential of the diffeomorphism defining the symmetry. 
\end{definition}
\begin{assumption}\label{assumption:G-invarinat neighborhood}
    The topology of the graph representation is G-invariant, i.e. if $V
_{\hat{\mathcal{G}}_t} \mapsto \mathcal{E}$ then $\phi_g(V
_{\hat{\mathcal{G}}_t}) \mapsto \mathcal{E},\forall g \in G$.
\end{assumption}
Assumption \ref{assumption:G-invarinat neighborhood} is important, as it ensures that if the state of the system is transformed via a group action, the safety of robot $i$ will still depend upon the same other agents and objects, whose state was transformed similarly to $i$'s. 
\begin{definition}
    Given a group $G$, a set $Y$ and group action $\rho:Y\times G \rightarrow Y$, the set $Y$ is G-invariant if $\forall y \in Y$ it holds that $\rho_g(y)\in Y,\forall g \in G$.
\end{definition}
\subsection{Exploiting Symmetries for Safe Policy Learning}\label{section:symmetries_cbf}
Let the safety set $S^r_{N_,i}:= \{ x^{{\mathcal{N}^{\hat{R}}_i}}\in \mathcal{X}^{|\mathcal{N}_i|+1} | \, c(x_i,x_j) \geq 0, \forall j \in \mathcal{E}_{\hat{\mathcal{G}}^i_t} \setminus \mathcal{E}_{*}\}  \,,\,\forall i \in I_N$ be a $G$-invariant set. If the specification $c:\mathcal{X}\times \mathcal{X}\rightarrow \mathbb{R}$ is $G$-invariant, i.e. $c(\phi_g(\alpha),\phi_g(\beta))=c(\alpha,\beta),\forall \alpha,\beta \in \mathcal{X},g\in G$ then $S_N$ is $G$-invariant (e.g. the Euclidean distance specification from Section \ref{intro:problem_statement} is $SE(n)$-invariant and its subgroups). 


\begin{definition}\label{definition_invariant_cbf}
    Consider a group $G$ and $G$-equivariant dynamics of Definition \ref{def:equiv_dynamics}. Consider a continuously differentiable, $G$-invariant function $h:\mathcal{X}^{|\mathcal{N}|+1}\rightarrow \mathbb{R}$ and the graph representation $\hat{\mathcal{G}}_t$ of section \ref{intro:graph_representation}. Let Assumptions \ref{assumption:G-invarinat neighborhood},\ref{ass:2} hold and $C_N \subset S^r_N$ with $C_N,S^r_N$ being $G$-invariant sets. Then $h$ is a valid $G$-invariant decentralized CBF if there exists a locally Lipschitz continuous extended class-$\mathcal{K}$ function $\alpha:\mathbb{R}\rightarrow \mathbb{R}$ such that the constraint of Equation \ref{cbf_contraint_def} holds.
\end{definition}
\begin{lemma}\label{lemma_ginvariance}
    For a $G$-invariant CBF of Definition \ref{definition_invariant_cbf}, the safety constraint of Equation \ref{cbf_contraint_def} is preserved under actions induced by elements of the group $G$.
\end{lemma}
\begin{proof}
    For clarity, denote $Q =\mathcal{N}^{\hat{R}}_i\cup \{i\}$ and let $g\in G$. Since $h$ is $G$-invariant, $h(x^{Q}) = h(\oplus_{k\in Q}\phi_g(x_k))$. Differentiating both sides w.r.t. time and leveraging the $G$-equivariance of the dynamics yields
    \begin{align*}
        \sum&_{j\in Q } \langle \nabla_{x_j} h(x^{Q}) , f(x_j,u_j)\rangle  = \dot{h}(x^Q) =  \\
        =\dot{h}&(\oplus_{k\in Q}\phi_g(x_k))=\sum_{j\in Q} \langle \nabla_{\phi_g(x_j)} h(\oplus_{k\in Q}\phi_g(x_k)), \frac{\partial}{\partial t} \phi_g(x_j) \rangle \\
        =& \sum_{j\in Q} \langle \nabla_{\phi_g(x_j)} h(\oplus_{k\in Q}\phi_g(x_k)), d\phi_g f(x_j,u_j) \rangle 
    \end{align*}
    Since $h$ is a valid CBF it holds that $\dot{h}(x^Q) \geq - \alpha(h(x^Q))$, yielding for $G$-invariance of $h$ and $G$-equivariant dynamics
    \begin{align}\label{constraint_}
        \sum_{j\in Q} \langle \nabla_{\phi_g(x_j)} h(\oplus_{k\in Q}\phi_g(x_k)), f(\phi_g(x_j),\psi_g(u_j)) \rangle \nonumber\\
        \geq -\alpha(h(\oplus_{k\in Q}\phi_g(x_k)))\,,\, \forall g \in G
    \end{align}\vspace{-7pt}
\end{proof}
Lemma \ref{lemma_ginvariance}, under the assumptions of Definition \ref{definition_invariant_cbf}, ensures that Lemma \ref{lemma:cbf} holds and forward invariance in the safe set $C_N$ is guaranteed.
\begin{lemma}\label{lemma_ginvariance_existence}
    For a $G$-equivariant dynamics $f:\mathcal{X}\times \mathcal{U}\rightarrow \mathfrak{X}$ and valid non-$G$-invariant CBF $h:\mathcal{X}\rightarrow \mathbb{R}$ on the $G$-invariant set $C:=\{x \in \mathcal{X}|\,h(x) \geq 0 \}$, there always exists an equivalent $G$-invariant valid CBF $\hat{h}:\mathcal{X}\rightarrow \mathbb{R}$.
\end{lemma}
\begin{proof}
    Consider the candidate CBF $\hat{h}(x)=\frac{1}{|G|}\int_{G}h(\phi_g(x))d\mu(g)$, where $\mu$ is the Haar measure. The function $\hat{h}$ is by construction $G$-invariant, owing to the $G$-invariance of the Haar measure. Since $C$ is $G$-invariant i.e. $\forall x\in C \Rightarrow \phi_g(x)\in C$, then $\forall x\in C$ it holds that $\hat{h}(x)\geq 0$ and $\dot{h}(\phi_g(x))\geq -\alpha(h(\phi_g(x))),\forall g\in G \Rightarrow\int_{G} \dot{h}(\phi_g(x))d\mu(g) \geq - \int_{G} \alpha(h(\phi_g(x)))d\mu(g)  $. Via the machinary of Lemma \ref{lemma_ginvariance}
    $\dot{\hat{h}}(x)= \frac{1}{|G|}\int_{G} \dot{h}(x)d\mu(g)$. As the integral over the group preserves the properties of the continuous extended class-$\mathcal{K}$ function, $\int_{G} \alpha(h(\phi_g(x)))d\mu(g)$ is also and continuous extended class-$\mathcal{K}$ function of $\hat{h}(x)$. Therefore $\dot{\hat{h}}(x) \geq -\bar{\alpha}(\hat{h}(x))$ and, thus, $\hat{h}$ is a valid CBF certifying forward invariance in $C$. This proof and can extended for the the multi-agent CBF terminology.
\end{proof}
\begin{figure}[hbt!]
    \centering
    \includegraphics[width=\linewidth]{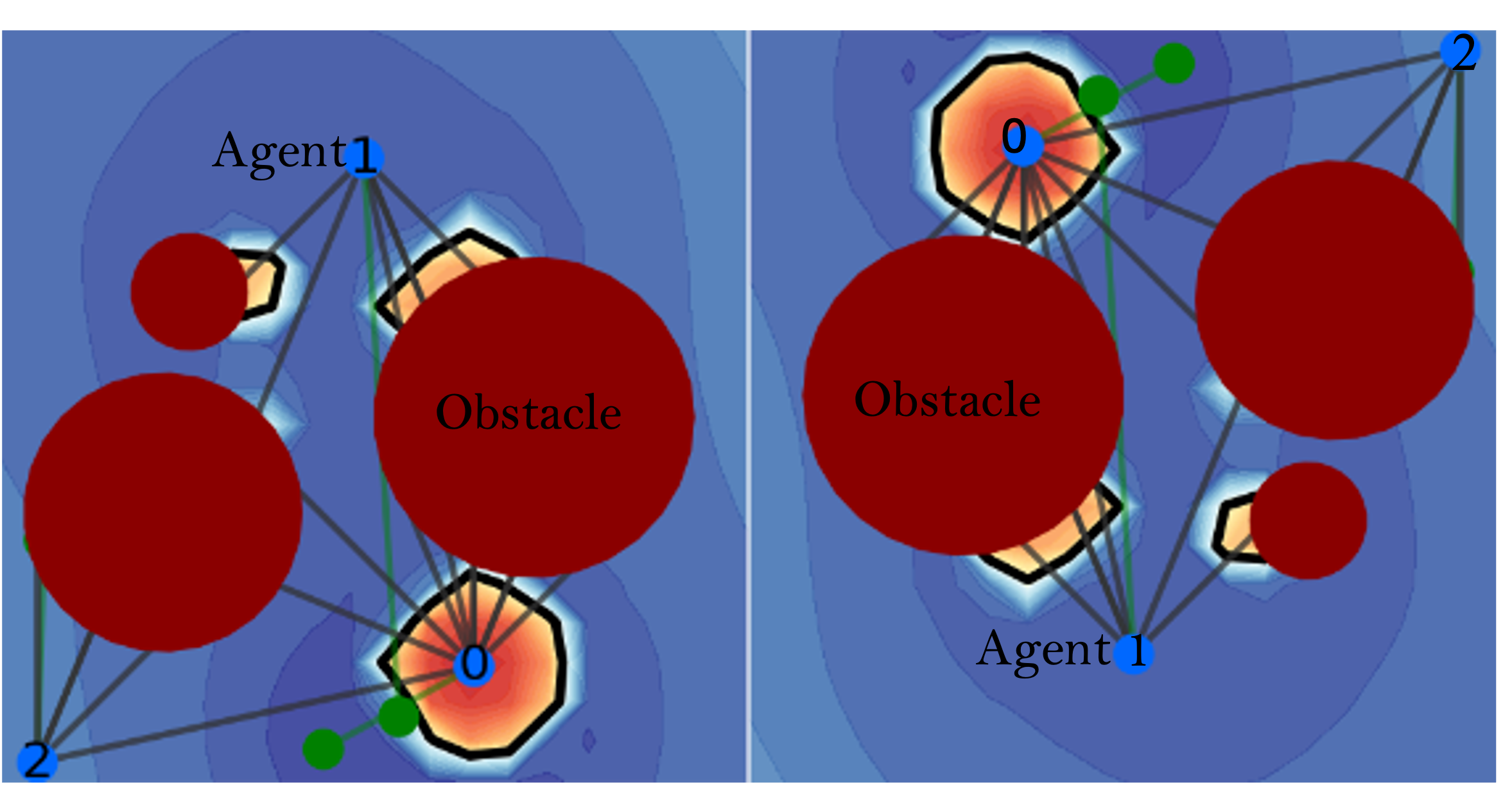}
    \caption{Top down views of Equivariant Neural Graph CBF contours from perspective of agent 1 for $V
_{\hat{\mathcal{G}}_t}$ and $\phi_g(V
_{\hat{\mathcal{G}}_t}), g\in SE(2)\times \mathbb{R}$. The topology of the graphs remains unchanged. Agent 2 is not a neighbor of agent 1 and does not affect the CBF (see Assumption \ref{assumption:G-invarinat neighborhood}).}
    \label{fig:equivcbf}\vspace{-10pt}
\end{figure}

\subsubsection{Why learn $G$-invariant CBFs? }
Without any loss of generality, consider the centralized MAS $x\in \mathcal{X}^N,u \in \mathcal{U}^N$. The following logic naturally extends to multi-agent systems via the constructions of Section \ref{section:graph_cbf}. Let $S_N$ be a $G$-set and assume $G$-equivariant dynamics. Assume that there exist a $G$-invariant valid CBF $h$ and a valid CBF $\hat{h}$ that does not exhibit symmetries in $G$, and let $C_N=\{x\in \mathcal{X}^N|h(x) \geq 0\},\hat{C}_N=\{x\in \mathcal{X}^N|\hat{h}(x) \geq 0\}$, thus $C_N$ is a $G$-invariant. As both $h,\hat{h}$ are valid CBFs, $\hat{C}_N\subset S_N, C_N\subset S_N$. Due to the lack of symmetries of $\hat{h}$, $\exists x_1,x_2\in S_N$ with $x_2=\phi_g(x_1)$ such that $x_1 \in \hat{C}_N,x_2 \notin \hat{C}_N$. But if $x_1\in C_N$, then by definition $x_2\in C_N$. As the safety constraint of equation \ref{cbf_constraint} stems from the forward invariance in $C_N$ and $\hat{C}_N$ respectively, of Lemma \ref{lemma:cbf}, then if $C_N\supset \hat{C}_N$ the constraints would be less restrictive for the $G$-invariant CBF, allowing greater flexibility for the controller. Increased coverage of $S_N$ by $C_N$ would lead to increased performance.
Additionally, shrinking the hypothesis class of the learned CBF to that of all G-invariant functions generally translates to fewer parameters (via parameter sharing \cite{bekkers2024fast}) required, faster convergence, and greater sampling efficiency as multiple $(x,u)$ pairs, identified via the induced actions of the group $G$, are essentially safety-equivalent. Therefore, for $G$-invariant $S_N$ and $G$-equivariant dynamics, any G-invariant CBF trained on a dataset $D=\{((x^1,u^1),(x^2,u^2),\dots \}$ would generalize to the dataset $\hat{D} = \cup_{\forall g\in G} \{(\phi_g(x^1),\psi_g(u^1)),(\phi_g(x^2),\psi_g(u^2)),\dots\}$ where $D\subseteq \hat{D} \subseteq \mathcal{X}^N$. Leveraging symmetries is particularly important in multi-agent systems as learning policies and safety certificate for them is sampling inefficient. We demonstrate experimentally in Section \ref{sec:experiments} the benefit of leveraging $G$-invariant models to learn multi-agent CBFs.


\subsubsection{Policy equivariance under invariant CBF constraints}
The optimal nominal controller $u^{*}_{i,\text{nom}}=\pi_\text{nom}(x_i)$ introduced as a surrogate in Section \ref{section:safe_distributed_policy} is the solution of
\begin{align}\label{eq:problem_Gnomialpolicy}
    u^{*}_{i,\text{nom}} =& \argmin_{u_i^{0:T}\in \mathcal{U}} \int_{0}^T \mathcal{T}(x_i(\tau)|\hat{x}_i ) d\tau \nonumber\\
    s.t.\quad& \dot{x}_i(t)=f(x_i(t),u_i(t))\in \mathfrak{X}(\mathcal{X})
\end{align}
\begin{theorem}\label{theorem:equivariant_optimal_control}
    Consider a group $G$ and transitive actions induced on the manifold $\mathcal{X}$, $\phi:\mathcal{X}\times G \rightarrow \mathcal{X},\psi:\mathcal{U}\times G \rightarrow \mathcal{U}$. For $G$-equivariant dynamics (Definition \ref{def:equiv_dynamics}) and $G$-invariant cost function $\mathcal{T}:\mathcal{X}\times \mathcal{X}\rightarrow \mathbb{R}$, the optimal nominal policy $\pi^{*}(x_i(t))$ for the problem of Equation \ref{eq:problem_Gnomialpolicy} is $G$-equivariant.
\end{theorem}
\begin{proof}
From the $G$-invariance of $\mathcal{T}$, for the value function it stands that $V_{t,\pi} (x(t)) = \int_{0}^{t} \mathcal{T}(x(\tau),\hat{x}) d\tau = V_{t,\psi_g(\pi)}(\phi_g(x(t)),\hat{x}),\, \forall g \in G$. For optimal value function $V^{*}$ acquired via the optimal policy $u^{*}(t)= \pi^{*}(x(t),\hat{x})$, from the Hamilton-Jacobi-Bellman equation $-V^{*}_t(x(t)) = \min_{u\in 
{\mathcal{U}}} \big[\mathcal{T}(x(t),\hat{x})+ \dot{V}^{*}_t(x(t))\big]$, we have $\pi^{*}(x(t)) = \arg\min_{u\in \mathcal{U}} \big[\mathcal{T}(x(t),\hat{x})+  \langle \nabla_{x(t)} V^{*}_t(x(t)), f(x(t),u(t)) \rangle \big]$. Differentiating the $G$-invariance constraint of the value function w.r.t. time and leveraging the $G$-equivariance of the dynamics, yields $\langle \nabla_{x(t)} V^{*}_t(x(t)), f(x(t),u(t)) \rangle = \nabla_{\phi_g(x(t))} V^{*}_t(\phi_g(x(t))), f(\phi_g(x(t)),\psi_g(u(t))) \rangle$. Therefore:
\begin{align*}
    &\pi^{*}(\phi_g(x(t))) = \arg\min_{\Tilde{u}\in \mathcal{U}} \big[\mathcal{T}(\phi_g(x(t)),\phi_g(\hat{x})) \\
    &\qquad\qquad +\langle \nabla_{\phi_g(x(t))} V^{*}_t(\phi_g(x(t))), f(\phi_g(x(t)),\Tilde{u}) \rangle \big] \\
    =& \arg\min_{\Tilde{u}\in \mathcal{U}} \big[\mathcal{T}(x(t),\hat{x}) 
     +\langle \nabla_{x(t)} V^{*}_t(x(t)), f(x(t),\psi_{g^{-1}}(\Tilde{u})) \rangle\big]\\
    &\overset{\Tilde{u}:=\psi_g(u)}= \arg\min_{\psi_g(u)\in \mathcal{U}} \big[\mathcal{T}(x,\hat{x}) +\langle \nabla_{x} V^{*}_t(x, f(x,u) \rangle\big]\\
    &= \psi_g(\pi^{*}(x(t)))\,,\, \forall g \in G
\end{align*}\vspace{-8pt}
\end{proof}
\begin{theorem}\label{theorem_equiv_policy}
    Consider $G$-equivariant dynamics (Def. \ref{def:equiv_dynamics}), $G$-invariant cost function $\mathcal{T}:\mathcal{X}\times \mathcal{X}\rightarrow \mathbb{R}$ and valid $G$-invariant CBF (Def. \ref{definition_invariant_cbf}). Then, the optimal policy for the safety constrained problem of Equation \ref{problem:3} is $G$-equivariant.
\end{theorem}
\begin{proof}
    If $u^Q=\pi(x^{Q})$ satisfies the original CBF constraints \ref{cbf_constraint}, then from Lemma \ref{lemma_ginvariance}, then $\forall \bar{x}^Q\in \mathcal{O}^{\phi}_{x^Q}$ the feasibility of the constraints is preserved for $\psi_g(\pi(x^Q))$ with $g$ s.t. $x^Q=\oplus_{k \in Q}\phi_{g^{-1}}(\bar{x}_k)$. Consider the optimization problem for $\bar{x}^Q$, $\pi^{*}(\bar{x}^Q) = \argmin_{u^Q} ||u^Q-\pi_\text{nom}(\bar{x}^Q)||$ subject to $\sum_{j\in Q} \langle \nabla_{\phi_g({x}_j)} h(\oplus_{k\in Q}\phi_g({x}_k)), f(\phi_g({x}_j),u_j) \rangle \geq -\alpha(h(\oplus_{k\in Q}\phi_g({x}_k)))\,,\, \forall g \in G$. Let $v^Q=\oplus_{k \in Q}\psi_{g^{-1}}(u_k)$. As Theorem \ref{theorem:equivariant_optimal_control} holds and the nominal controller is $G$-equivariant, i.e. $\pi_\text{nom}(\oplus_{k \in Q}\phi_g(x_k))=\oplus_{k \in Q}\psi_g\pi_\text{nom}(x^Q)$, and the group actions preserve norms, then $\pi^{*}(\oplus_{k \in Q}\phi_{g}(x_k)) = \oplus_{k \in Q}\psi_{g}(\argmin_{v^Q} ||\oplus_{k \in Q}\psi_{g}\pi(x^Q)- \oplus_{k \in Q}\psi_{g}\pi_\text{nom}(x^Q)||)_k = \oplus_{k \in Q}\psi_{g}(\argmin_{v^Q} ||v^Q- \pi_\text{nom}(x^Q)||)_k$ subject to $\sum_{j\in Q } \langle \nabla_{x_j} h(x^{Q}) , f(x_j,u_j)\rangle$ by the constraint preservation under actions from $G$. But this is the original problem for $x^Q$ with solution $\pi^{*}({x}^Q)$. Thus, we conclude that $\pi^{*}(\oplus_{k \in Q}\phi_g(x_k))=\oplus_{k \in Q}\psi_g\pi^{*}(x^Q)_k$
\end{proof}

For $G$-equivariant dynamics, $G$-invariant CBF and $G$-invariant cost function $\mathcal{T}$, Theorem \ref{theorem:equivariant_optimal_control}, shows that the surrogate nominal controller $u_\text{nom}$ must be a $G$-equivariant function. Using $u_\text{nom}$ we then aquire $G$-equivariant (Theorem \ref{theorem_equiv_policy}) $u_\text{QP}$ that the loss function incentivizes $\pi_\theta$ to learn to stay as close as possible. Parametrizing the learnable policy $\pi_\theta$ by a $G$-equivariant neural network offers significant generalization banefits and boosts sampling efficiency \cite{chen2024rme3equivariantactorcriticmethods}.

\subsection{Equivariant Graphormer}\label{section:LieGraphormer}
To solve the problem of Section \ref{section:safe_distributed_policy} with the structure of Section \ref{section:symmetries_cbf}, we must learn the G-equivariant policy $\pi_{\theta}(\hat{\mathcal{G}}_t^{i})$ and G-invariant decentralized CBF $h_{\phi}(\hat{\mathcal{G}}_t^{i})$. Generally, deep learning models require specialized architectures to ensure satisfaction of the equivariant constraints (e.g. \cite{fuchs2020se3transformers}), resulting in challenging optimization, slow inference and limited of adaptability to new manifolds. To address these challenges we leverage the structure of our graph representation to achieve equivariance via group actions, without additional constrains, thus, allowing for easy extension of pre-existing architectures, similarly to \cite{lippmann2025beyond,bousias2025symmetriesenhancedmultiagentreinforcementlearning}. 

\subsubsection{Group Canonicalization}\label{sec:groupcanon}
Given a group $G$ acting on space $X$ via $\phi_g$, we can define an extended group action $\phi'_g$ on space $G\times X$ as $\phi'_g[(p,x)]=(g\cdot p,\phi_g[x])$, $\forall g,p\in G, x\in X$. Since $\phi_g$ is an action of $G$, $\phi'_g$ satisfies the properties of definition \ref{definitioin:group_action} and, thus, it is also an action.
Consider a space $Y$ with corresponding group action $\psi_g:G\times Y\to Y$.
\begin{lemma}\label{lem:canonic}
A function $f:G\times X\to Y$ satisfies the equivariant constraint $f(\phi'_g[p])=\psi_g[f(p)]$, $\forall g\in G$ and $p\in G\times X$, if and only if, for $h:X\to Y$, it can be written as a composition $f(g,x)=\psi_g[h(\phi_{g^{-1}}[x])],\quad \forall g\in G, x\in X$
with $h(x)=f(e,x)$ for all $x\in X$ and $e$ being the identity element of $G$.
\end{lemma}
Notice that in Lemma \ref{lem:canonic}, $h$ is a function from $X$ to $Y$ without any additional constraints. This implies that we can use any general function approximator (e.g. MLP, Transformer) to approximate an equivariant function $f:G\times X\to Y$, by only applying the appropriate group transformations. We will use this observation to extend the baseline non-equivariant models to be equivariant with minimal changes to the underlying architecture.
\subsubsection{Equivariant Graph Transformer}\label{sec:equivgraphTransformer}
 Given a feature augmented graph representation $(V,E,F)$ with a finite set of nodes $V$, a finite set of edges $E(G)\subset \{(u,v)|u,v\in V\}$ and a set of per-node features $F=\{f_v\in X|v\in V\}$, a graph transformer sequentially updates the nodes features using a local attention layer to aggregate information from neighboring nodes. Specifically the $l^{th}$ update layer for node $v\in V$ is a function $M_v: X^{(l)}\to X^{(l+1)}$ defined as $f_v^{(l+1)}\leftarrow M_v(F^{(l)})=\mu(f_v^{(l)}+A(F^{(l)})_v)$, where $\mu$ corresponds to a fully-connected feedforward network, and $\mathrm{attn}$ corresponds to the local attention layer: 
\begin{align*}
    A(F^{(l)})_v=\sum_{p\in \mathcal{N}_v} \text{softmax}\left( {f_v^{(l)}}^T W_Q^T W_Kf_p^{(l)} \right)\left(W_V f_p^{(l)} \right)
\end{align*}
with $\mathcal{N}_v$ being the set of neighbors for nodes $v$.
As discussed in Section \ref{intro:graph_representation}, the input graph representation is endowed an additional structure that allows for an simple extension of the graph transformer used in the baselines to become equivariant. Specifically, each node $v\in V$ additional to a feature $f_v\in X$ describing each state is also equipped with a local frame $g_v\in G$. This means that the input graph is represented  as $(V,E,F_\mathrm{tens})$, with $F_\mathrm{tens}=\{(g_v,f_v)\in G\times X|v\in V\}$ being a set of "tensorial" features that are described by their local frame $g_v\in G$ along with their feature value $f_v$. For such a feature $(g_v,f_v)$ expressed in local frame $g_v\in G$ we can compute the equivalent feature expressed in a new frame $g_n\in G$ by applying the action $\phi'_{g_ng_v^{-1}}[(g_v,f_v)]=(g_n,\phi_{g_ng_v^{-1}}[f_v])$. This structure allows us to leverage the results of Lemma \ref{lem:canonic} and define an equivariant update rule $M^\mathrm{eq}_v:G\times X^{(l)}\to X^{(l+1)} $ for the features of node $v\in V$ as follows:
\vspace{-2mm}\begin{align*}
    f_v^{(l+1)}\leftarrow M^\mathrm{eq}_v(g_v,F^{(l)})=\phi^{(l+1)}_{g_v}\left[M_v\left(\phi^{(l)}_{g_v^{-1}}\left[F^{(l)}\right]\right)\right]
\vspace{-2mm}\end{align*}
with $\phi^{(l)}, \phi^{(l+1)}$ being actions of group $G$ on the corresponding input/ouput feature space $X^{(l)}, X^{(l+1)}$. By sequentially composing the equivariant update layers we define an edge-preserving isomorphic, permutation equivariant, end-to-end G-equivariant GNN, which is used to learn $\pi_{\theta},h_{\phi}$.

\section{Experiments}\label{sec:experiments}

\begin{table*}[t!]
  \centerfloat
\begin{adjustbox}{width=1.2\textwidth}
\begin{tabular}{c}
     \textsc{Swarm size (Agent Density $N/l^3$)}\\

\begin{tabular}{c|ccccc|ccccc|ccccc|} 
    \cline{2-16}
    & & & \textbf{8 (0.125)} & & & & & \textbf{16 (0.25)} & & & & & \textbf{64 (1)} & &   \\
    \cline{2-16}
        & Safe (\%)$\uparrow$ & Reach (\%)$\uparrow$ & Succ (\%)$\uparrow$ & Cost $\downarrow$& Rew $\uparrow$ & Safe (\%)$\uparrow$ & Reach (\%)$\uparrow$ & Succ (\%)$\uparrow$ & Cost $\downarrow$& Rew $\uparrow$ & Safe (\%)$\uparrow$ & Reach (\%)$\uparrow$ & Succ (\%)$\uparrow$ & Cost $\downarrow$& Rew $\uparrow$\\
    \hline\hline
    GCBF   & \textbf{100} & \textbf{100} & \textbf{100}& \textbf{0} (0/0) & -315.1
           & 97.5 & \textbf{100} & 97.5& 0.33 (0/2) & -311.4
           & 87.7& 99.8& 87.5& 1.09 (0.31/1.96) & -315\\
    GCBF+  & \textbf{100} & \textbf{100} & \textbf{100} & \textbf{0} (0/0) & -17.1
           & \textbf{100} & \textbf{100} & \textbf{100} & \textbf{0} (0/0) & -17.5
           & 96.5 & \textbf{100} & 96.5 & 0.23 (0/1) & -49.1\\
    dCBF &  \textbf{100}  & \textbf{100} & \textbf{100} & \textbf{0} (0/0) & -96.3
          &   \textbf{100}  & 98.125 & 98.13 & \textbf{0} (0/0) & -182.8
            &  -  & - & - & - & -\\
    cCBF &  \textbf{100}  & \textbf{100} & \textbf{100} & \textbf{0} (0/0) & -60.8
                  &   \textbf{100}  & \textbf{100} & \textbf{100} & 0 (0/0) & -120.5
                   &  -  & - & - & -  & -\\
    EGCBF(Ours)   & \textbf{100} & 91.2 & 91.2 & \textbf{0} (0/0) & -349.9
                  & \textbf{100} & 93.7 & 93.7 & \textbf{0} (0/0) & -338.5
                  & 88.4 & 94.2 & 83.4 &  1.256 (0.25/2) & -335.8\\
    EGCBF+(Ours)  & \textbf{100} & \textbf{100} &\textbf{100}& \textbf{0} (0/0) &\textbf{-9.3}
                  & \textbf{100} & \textbf{100} & \textbf{100} &  \textbf{0} (0/0) &\textbf{-14.1}
                  & \textbf{98.8} & 99.7 & \textbf{98.5}& \textbf{0.04} (0/0.21) &\textbf{-34.7}\\
    \hline

    \cline{2-16}
    & & & \textbf{128 (2)} & & & & & \textbf{256 (4)} & & & & & \textbf{512 (8)} & &   \\
    \cline{2-16}
        & Safe (\%)$\uparrow$ & Reach (\%)$\uparrow$ & Succ (\%)$\uparrow$ & Cost $\downarrow$& Rew $\uparrow$ & Safe (\%)$\uparrow$ & Reach (\%)$\uparrow$ & Succ (\%)$\uparrow$ & Cost $\downarrow$& Rew $\uparrow$ & Safe (\%)$\uparrow$ & Reach (\%)$\uparrow$ & Succ (\%)$\uparrow$ & Cost $\downarrow$& Rew $\uparrow$\\
    \hline\hline
    GCBF   & 73.9 & 99.1& 73.2& 2.47 (1.72/4.12) & -350.5
           & 52.1 & 98.5& 51.1& 5.98 (4.66/7.46) & -409.3
           & 31.6 & 96.8& 30& 11.07 (9.97/13.4) & -505.8\\
    GCBF+  & 93.6 & 99.3 & 92.9& 0.43 (0/1.31) & -104.2
           & 88.8 & \textbf{98.7} & 87.7& 0.72 (0.56/0.92) & -209.9
           & 77.8 & 96.2 & 74.5& 1.51 (1.2/1.72) & -484.2\\
    EGCBF(Ours)   & 74.1 & 96.7 & 71.7 & 2.9(2.04/3.79) & -358.1
                  & 55.1 & 97.5 & 53.6 & 5.42(4.23/7.36) & -409.8
                  & 33.1 & \textbf{98.7} & 31.7 & 10.1 (7.03/15.29)& -487.3\\
    EGCBF+(Ours)  & \textbf{97.9} &\textbf{99.5}& \textbf{97}& \textbf{0.12} (0/0.23) &  \textbf{-52.1}
                  & \textbf{94.2} & 96.6& \textbf{91.1} & \textbf{0.26} (0.06/0.51)  &\textbf{-97.8} 
                  & \textbf{89.5} & 92.1& \textbf{83.3}& \textbf{0.51} (0.23/0.73) & \textbf{-224.8} \\
    \hline    
\end{tabular}

\end{tabular}
  \end{adjustbox}
\caption{Evaluation of zero-shot transferability to growing swarm sizes and agents density for $l=4$ without obstacles. Best in class shown in bold.}
\label{table:results_scalability}
\end{table*}

We offer simulation experiments to evaluate the effect on scalability and generalization of incorporating geometric symmetries in the Graph-based CBF and policy models. 

\noindent\textbf{Environment}:
We consider a swarm of $N$ quadrotors:
\begin{align*}
    \ddot{p}_i =& \mathbf{g} + \frac{1}{m}R_i \mathbf{F}_i \\
    \dot{\omega}_i =& J^{-1}(\tau_i - \omega_i \times (J \omega_i)) \\
    \dot{R}_i =& \omega_{i,\times} R_i
\end{align*}
where $i\in I_N$ the agent tag, $m\in \mathbb{R}_{+}$ denotes the mass, $\ddot{p}_i \in \mathbb{R}^3$ is the acceleration in the world frame, $\mathbf{g}=[0,0,-9.81]^T$ is the gravitational acceleration vector, $R_i\in SO(3)$ is the rotation matrix from the body frame to the world frame, $F_i = F_3\mathbf{e}_3 \in\mathbb{R}^3,F_3\in \mathbb{R}^+$ is the thrust force in the body frame, $\omega_i \in \mathbb{R}^3$ is the angular velocity in the body frame, $J\in \mathbb{R}^{3 \times 3}$ is the inertia tensor, $\tau_i\in \mathbb{R}^3$ is the torque in the body frame for some function $\kappa$ that computes the torques induced in the local frame by the motor speeds, and $\omega_{i,\times} \in \mathbb{R}^{3 \times 3}$ is a skew-symmetric matrix associated with the angular acceleration rotated to the world frame. 
The dynamics can be compactly written in a control affine system as $\dot{x}_i(t) = f_0(x_i(t))+ f_1(x_i(t))u_i(t))$, where the state comprises of pose $\rho_i=(p_i,R_i)$ and twist $\xi_i=(\dot{p}_i,\omega_i)$, i.e. $x_i(t) = [\rho_i(t),\xi_i(t)] \in SE(3) \times \mathbb{R}^6$ and $u_i(t) = [\tau_i, F_3]\in \mathbb{R}^3 \times \mathbb{R}$.  As gravity is $O(3)$-invariant, the $SE(3)$-symmetry is broken, but the system is equivariant with respect to the subgroup of $SE(3)$
\begin{align}
    \bar{G} = \Big\{\begin{pmatrix} R^z(\theta) & \lambda  \\ 0 & 1 
    \end{pmatrix} | (\lambda,\theta)\in \mathbb{R}^3 \times S^1 \Big\}
\end{align}
which is a a Lie group, isomorphic to $SE(2) \times \mathbb{R}$, acting on $SE(3) \times \mathbb{R}^6$ via $\phi_g(x)=(g\rho,\xi)$ and on $\mathbb{R}^3 \times \mathbb{R}$ via $\psi_g(u)=(g \tau, F_3)$. If the neighborhood $\mathcal{N}_i$ is constructed via a Euclidean relative distance, for the $\bar{G}$ group induced actions Assumption \ref{assumption:G-invarinat neighborhood} stands. 

\noindent\textbf{Data collection}: In all episodes initial, target and obstacle positions are uniformly sampled over $[0,l]^3$. The training dataset $\mathcal{D}=\{D_{\text{safe},i},D_{\text{unsafe},i}\,\forall i \in I_N\}$ is collected over numerous random configurations via an on-policy strategy that periodically executes the learned policy $\pi_{\theta}(\hat{G}_t)$ that limits distributional shift between training and inference data. Labelling the data via the distance-based safety specification may lead to infeasibility issues and computing the infinite horizon control invariant set is computationally intractable. Thus, to partition the dataset, we employ the finite reachibility approximation of \cite{zhang2025gcbf+}. If $\hat{\mathcal{{G}}}^i_t$ entails collisions, it is added in $D_{\text{unsafe},i}$. If not, $\pi_{\theta}(\hat{G}_t)$ is used to unroll sampled trajectories for a fixed horizon $T$ and, if no future collision is detected, the state is added in $D_{\text{safe},i}$, otherwise it remains unlabeled. Setting $T\rightarrow \infty$, this process would recover the infinite horizon control invariant set. The construction of graph representation of the state of the system $\hat{\mathcal{G}}_t$ is described in Section \ref{intro:graph_representation}. Notice that the torsor $\mathfrak{B}_{\bar{G}}$ of the group $\bar{G}$ is a submanifold of the state, thus, allowing for a group representation element to form from the pose to construct the $\bar{G}$-augmented graph representation $\hat{\mathcal{G}}_t$. 

\noindent\textbf{Task \& Evaluation Metrics}:
In each training episode, every quadrotor attempts to reach a designated target position of the formation while avoiding collisions with obstacles and other quadrotors. As metrics, we report the \textit{mean safety rate} (\% of agents not in collision with objects or other agents throughout the episode over all agents), \textit{mean reach rate} (\% of agents that satisfy liveness condition of reaching the target), \textit{mean success rate} (\% of agents reaching their target while remaining collision-free throughout the episode), \textit{mean cost} (average number of collisions per episode per agent) and \textit{cumulative reward} (negative sum $L_2$ loss of policy and reference controller throughout the episode). We evaluated the performance of the models in 50 episodes per swarm size and density of agents ($\approx N/l^3$).

\subsection{Architecture \& Training}\label{section:architecture_training}
The learning framework consists of a $\bar{G}$-equivariant neural network for the policy $\pi_{\theta}(\hat{\mathcal{G}}^i_t) = \psi_{g_i} \circ m_{\theta_2}\circ \phi_{g_i^{-1}} \circ M_{\theta_1}(\hat{\mathcal{G}}^i_t)$ and a $\bar{G}$-invariant neural network for the distributed CBF $h_{\phi}(\hat{\mathcal{G}}^i_t) = m_{\phi_2}\circ \phi_{g_i^{-1}} \circ M_{\phi_1}(\hat{\mathcal{G}}^i_t)$, trained in parallel with the loss function of Section \ref{section:safe_distributed_policy}, where $M_{\theta_1},M_{\phi_1}$ are $\bar{G}$-equivariant graphormers of Section \ref{sec:equivgraphTransformer} and $m_{\theta_2},m_{\phi_2}$ standard MLPs. The structure of our networks is closely resembles \cite{zhang2025gcbf+}, with the addition of (de)canonicalizing group actions. As the torque and thrust were assumed to be in the local frame $\psi_{g_i^{-1}} \circ \pi_{\theta}(\hat{\mathcal{G}}^i_t)$. After the sampled $\hat{\mathcal{G}}^i_t$ is labeled safe/unsafe according to the aforementioned data collection scheme via $\pi_\theta$, $h_\phi,\pi_\text{nom}$ are used to solve the centralized QP-CBF of Section \ref{section:safe_distributed_policy} to produce $\pi_\text{QP}$. Finally, the errors from the loss function $\mathcal{L}$ are backpropagated to update the networks. We offer two variations of the $\bar{G}$-invariant CBF $h_\phi$: EGCBF+ and EGCBF that use $\pi_\text{QP}$ and $\pi_\text{nom}$ as reference controllers in the loss function. We use as baselines GCBF+ \cite{zhang2025gcbf+}, GCBF \cite{zhang2023gcbf} without the online policy refinement step, along with centralized and decentralized variants of CBF-QP \cite{7857061,zhang2025gcbf+}.  All models are trained with the Adam optimizer on an AMD EPYC 7313P-16 CPU @ 3000MHz and an NVIDIA A100 GPU, with $l=2$, learning rates $10^{-4}$ for the CBF and $10^{-5}$ for the policy, $\gamma=0.02$, $\eta_{\textbf{d}}=0.2$ and linear extended class-$\mathcal{K}$ function $\alpha(\cdot)=\cdot$. All other hyperparameters regarding the architecture and training algorithm follow \cite{zhang2025gcbf+}.
\begin{figure}[hbt!]
    \centering
    \includegraphics[trim={1cm 0.2cm 1cm 1.5cm},clip,width=\linewidth]{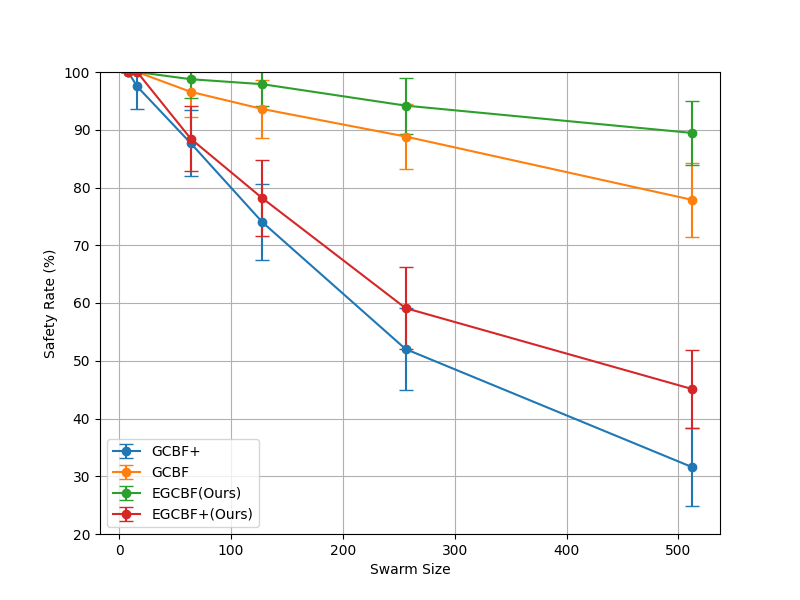}
    \caption{Safety rates for zero-shot scalability evaluations for $l=4$ and 5 obstacles. The safety rates are dropping only slightly compared to the obstacle-free experiments, suggesting that obstacle avoidance is easier for the static parts of the graph than with other agents (that requires coordination). Still, the symmetry-enhanced models outperform their non-symmetric counterparts.}
    \label{fig:scalability}\vspace{-15pt}
\end{figure}
\subsection{Generalization \& Scalability Analysis}
We offer experiments to evaluate the scalability, generalization and efficiency of the proposed symmetries-enhanced network. The models $\pi_\theta,h_\phi$ were trained in navigation scenarios with a swarm comprising of 8 robots, $l=2$ and 5 obstacles for 1000 steps, and tested with zero-shot transfer on swarms of 8-512 robots \footnotemark[\value{footnote}], with $l=4$ (estimated density ranging from 0.125 to 8, an increase of 6400\%), in environments without obstacles (summarized in Table \ref{table:results_scalability}) and with 8 obstacles (see Figure \ref{fig:scalability}). We were unable to test cCBF-QP and dCBF-QP for more than 16 robots as solving the QP exceeded computation time and memory restrictions (the dCBF-QP would theoretically run on multiple robots separately, and thus would run for more robots, but our simulation is run centralized on one node). For 8 robots, same as during training, all methods guarantee safety, though the hand-crafted cCBF-QP, dCBF-QP, and EGCBF, GCBF (trained with the nominal controller instead of the QP solution) are overly conservative, as indicated by the high reach rate but low reward. As $N$ increases, EGCBF and GCBF safety and success rates drop significantly (by $\approx 70\%$ and 60\% respectively), while the reach rate remains high (the low rewards suggest a slow target reaching with possible deadlocks). This can be explained by the fact that the nominal controller is computed without any notion of safety, thus the control and safety terms in the loss function are inconsistent. Figure \ref{fig:scalability} demonstrates the significance of the training the models with reference controllers that account for safety. The safety rates of (E)GCBF drop exponentially, contrary to (E)GCBF+ that is almost linear. However, as environments get more cluttered, (E)GCBF+ reach rates drop faster, as $\pi_\text{nom}$ that ensures liveness is unable to resolve deadlocks when adapted by the $\pi_{QP}$.   
Overall, the symmetry-enhanced EGCBF+ outperforms the baselines across all sizes-densities, attaining higher rewards and exhibiting to fewer collisions (for 512 drones the EGCBF+ collision rate drops to $\approx 90\%$ compared to $<78\%$ for the baselines) than the baselines and increased success rate, as the embedded symmetries that are leveraged locally are not affected by the size of the swarm. It is indicative that, for 512 drones, EGCBF+ leads to $0.5$ collisions per drone while the non-equivariant more than $1.5$.
\begin{figure}[hbt!]
    \centering
    \includegraphics[trim={2.4cm 2cm 2cm 3cm},clip,width=\linewidth]{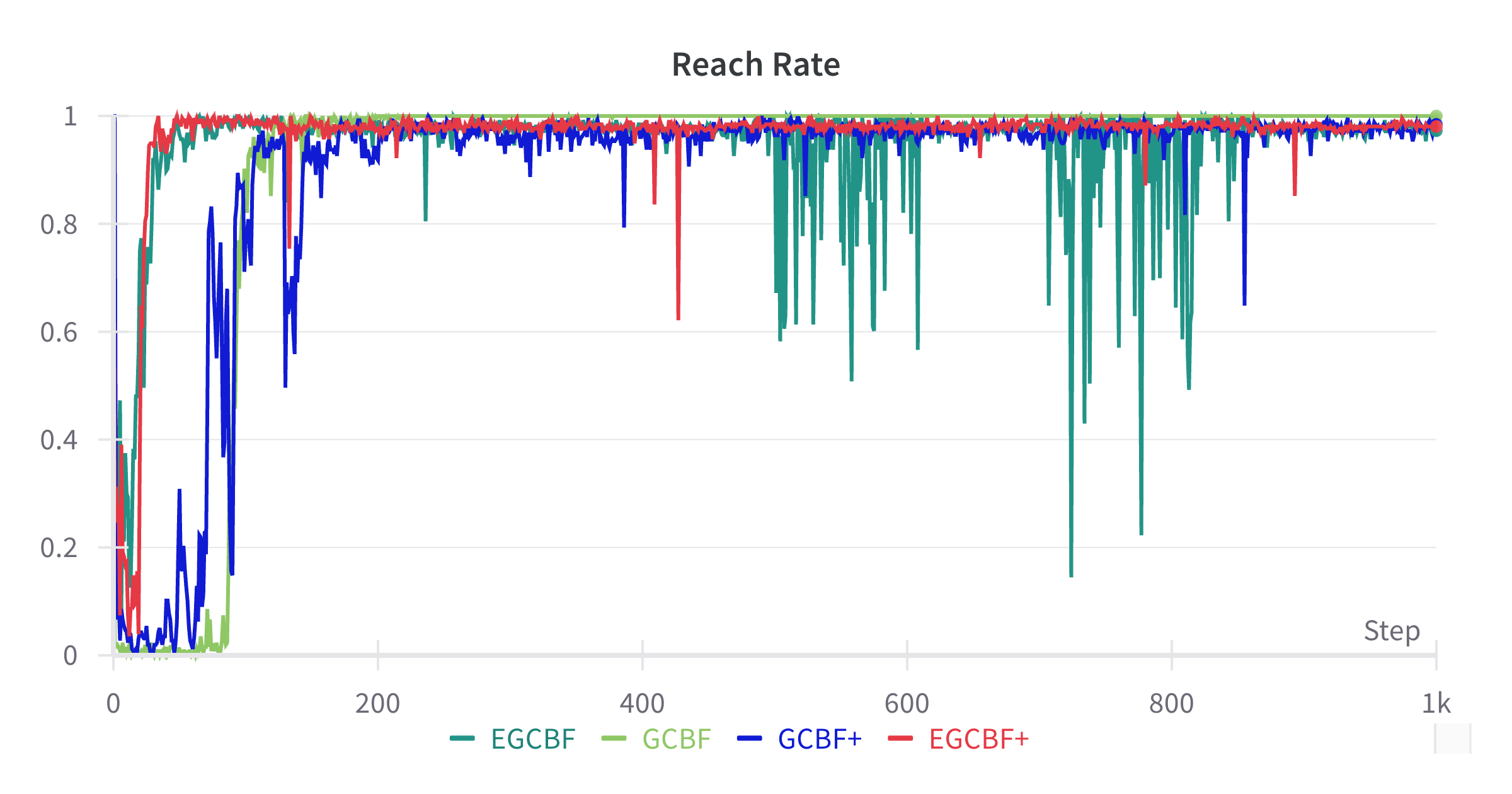}
    \caption{$G$-equivariant policies converge faster than the non-equivariant ones.}
    \label{convergence}
\end{figure}
The equivariant policies exploit the structure of the collision avoidance and liveness specification to shrink the hypothesis class and, thus, effectively restricting the models and leading to sampling efficient and accurate learning of collision avoidance
without loss of expressivity of the final policy. 
EGCBF(+) converge faster (Figure \ref{convergence} depicts reach rate during training) than GCBF(+) as the $G$-equivalent state-action pairs appear.

\section{Conclusions}
This paper motivates the embedding of symmetries in multi-agent graph-based Control Barrier Functions, and introduces a framework for learning symmetries-infused distributed safe policies via a group-modular equivariant neural network. The experimental results of this work suggest that embedding intrinsic symmetries in data-driven safety-constrained policies offers benefits in generalization, scalability and sample efficiency of the model. But what if the dynamics and the assumed safe set have different symmetries? We leave this question for future work.


\bibliography{main.bib}

\begin{thebibliography}{10}

\bibitem{10529204}
Z.~Dong, S.~Omidshafiei, and M.~Everett, ``Collision avoidance verification of multiagent systems with learned policies,'' {\em IEEE Control Systems Letters}, vol.~8, pp.~652--657, 2024.

\bibitem{7798509}
M.~Chen, J.~C. Shih, and C.~J. Tomlin, ``Multi-vehicle collision avoidance via hamilton-jacobi reachability and mixed integer programming,'' in {\em CDC}, pp.~1695--1700, 2016.

\bibitem{7911207}
Y.~Lin and S.~Saripalli, ``Sampling-based path planning for uav collision avoidance,'' {\em IEEE Transactions on Intelligent Transportation Systems}, vol.~18, no.~11, pp.~3179--3192, 2017.

\bibitem{8613928}
H.~Zhu and J.~Alonso-Mora, ``Chance-constrained collision avoidance for mavs in dynamic environments,'' {\em IEEE Robotics and Automation Letters}, vol.~4, no.~2, pp.~776--783, 2019.

\bibitem{7989037}
Y.~F. Chen, M.~Liu, M.~Everett, and J.~P. How, ``Decentralized non-communicating multiagent collision avoidance with deep reinforcement learning,'' in {\em ICRA}, pp.~285--292, 2017.

\bibitem{zhang2024scalable}
L.~Zhang, L.~Li, W.~Wei, H.~Song, Y.~Yang, and J.~Liang, ``Scalable constrained policy optimization for safe multi-agent reinforcement learning,'' in {\em NeurIPS}, 2024.

\bibitem{ames2016control}
A.~D. Ames, X.~Xu, J.~W. Grizzle, and P.~Tabuada, ``Control barrier function based quadratic programs for safety critical systems,'' {\em IEEE Transactions on Automatic Control}, vol.~62, 2016.

\bibitem{glotfelter2017nonsmooth}
P.~Glotfelter, J.~Cort{\'e}s, and M.~Egerstedt, ``Nonsmooth barrier functions with applications to multi-robot systems,'' {\em IEEE control systems letters}, vol.~1, no.~2, pp.~310--315, 2017.

\bibitem{9642050}
X.~Tan and D.~V. Dimarogonas, ``Distributed implementation of control barrier functions for multi-agent systems,'' {\em IEEE Control Systems Letters}, vol.~6, pp.~1879--1884, 2022.

\bibitem{8718798}
L.~Lindemann and D.~V. Dimarogonas, ``Control barrier functions for multi-agent systems under conflicting local signal temporal logic tasks,'' {\em IEEE Control Systems Letters}, vol.~3, no.~3, pp.~757--762, 2019.

\bibitem{9992744}
H.~Parwana, A.~Mustafa, and D.~Panagou, ``Trust-based rate-tunable control barrier functions for non-cooperative multi-agent systems,'' in {\em 2022 IEEE 61st Conference on Decision and Control (CDC)}, pp.~2222--2229, 2022.

\bibitem{10.1109/CDC42340.2020.9303785}
A.~Robey, H.~Hu, L.~Lindemann, H.~Zhang, D.~V. Dimarogonas, S.~Tu, and N.~Matni, ``Learning control barrier functions from expert demonstrations,'' in {\em CDC}, 2020.

\bibitem{mdcbf2021}
Z.~Qin, K.~Zhang, Y.~Chen, J.~Chen, and C.~Fan, ``Learning safe multi-agent control with decentralized neural barrier certificates,'' in {\em International Conference on Learning Representations}, 2021.

\bibitem{zhang2023gcbf}
S.~Zhang, K.~Garg, and C.~Fan, ``Neural graph control barrier functions guided distributed collision-avoidance multi-agent control,'' in {\em Conference on Robot Learning}, pp.~2373--2392, PMLR, 2023.

\bibitem{zhang2025gcbf+}
S.~Zhang, O.~So, K.~Garg, and C.~Fan, ``Gcbf+: A neural graph control barrier function framework for distributed safe multi-agent control,'' {\em IEEE Transactions on Robotics}, pp.~1--20, 2025.

\bibitem{gao2024provably}
Z.~Gao, G.~Yang, J.~Bayrooti, and A.~Prorok, ``Provably safe online multi-agent navigation in unknown environments,'' in {\em 8th Annual Conference on Robot Learning}, 2024.

\bibitem{bekkers2024fast}
E.~J. Bekkers, S.~Vadgama, R.~Hesselink, P.~A.~V. der Linden, and D.~W. Romero, ``Fast, expressive \${\textbackslash}mathrm\{{SE}\}(n)\$ equivariant networks through weight-sharing in position-orientation space,'' in {\em The Twelfth International Conference on Learning Representations}, 2024.

\bibitem{vanderpol2022multiagentmdphomomorphicnetworks}
E.~van~der Pol, H.~van Hoof, F.~A. Oliehoek, and M.~Welling, ``Multi-agent mdp homomorphic networks,'' 2022.

\bibitem{mcclellan2024boostingsampleefficiencygeneralization}
J.~McClellan, N.~Haghani, J.~Winder, F.~Huang, and P.~Tokekar, ``Boosting sample efficiency and generalization in multi-agent reinforcement learning via equivariance,'' 2024.

\bibitem{pmlr-v139-satorras21a}
V.~G. Satorras, E.~Hoogeboom, and M.~Welling, ``E(n) equivariant graph neural networks,'' in {\em ICML}, 2021.

\bibitem{fuchs2020se3transformers}
F.~B. Fuchs, D.~E. Worrall, V.~Fischer, and M.~Welling, ``Se(3)-transformers: 3d roto-translation equivariant attention networks,'' in {\em NeurIPS}, 2020.

\bibitem{INFOMARL}
S.~Nayak, K.~Choi, W.~Ding, S.~Dolan, K.~Gopalakrishnan, and H.~Balakrishnan, ``Scalable multi-agent reinforcement learning through intelligent information aggregation,'' ICML'23, 2023.

\bibitem{8796030}
A.~D. Ames, S.~Coogan, M.~Egerstedt, G.~Notomista, K.~Sreenath, and P.~Tabuada, ``Control barrier functions: Theory and applications,'' in {\em ECC 2019}, pp.~3420--3431, 2019.

\bibitem{XU201554}
X.~Xu, P.~Tabuada, J.~W. Grizzle, and A.~D. Ames, ``Robustness of control barrier functions for safety critical control,'' {\em IFAC-PapersOnLine}, vol.~48, pp.~54--61, 2015.
\newblock Analysis and Design of Hybrid Systems.

\bibitem{tzes2023graph}
M.~Tzes, N.~Bousias, E.~Chatzipantazis, and G.~J. Pappas, ``Graph neural networks for multi-robot active information acquisition,'' in {\em 2023 IEEE International Conference on Robotics and Automation (ICRA)}, pp.~3497--3503, IEEE, 2023.

\bibitem{chen2024rme3equivariantactorcriticmethods}
D.~Chen and Q.~Zhang, ``${\rm e}(3)$-equivariant actor-critic methods for cooperative multi-agent reinforcement learning,'' 2024.

\bibitem{lippmann2025beyond}
P.~Lippmann, G.~Gerhartz, R.~Remme, and F.~A. Hamprecht, ``Beyond canonicalization: How tensorial messages improve equivariant message passing,'' in {\em ICLR}, 2025.

\bibitem{bousias2025symmetriesenhancedmultiagentreinforcementlearning}
N.~Bousias, S.~Pertigkiozoglou, K.~Daniilidis, and G.~Pappas, ``Symmetries-enhanced multi-agent reinforcement learning,'' 2025.

\bibitem{7857061}
L.~Wang, A.~D. Ames, and M.~Egerstedt, ``Safety barrier certificates for collisions-free multirobot systems,'' {\em IEEE Transactions on Robotics}, vol.~33, no.~3, pp.~661--674, 2017.

\end{thebibliography}
\bibliographystyle{ieeetr}


\end{document}